\documentclass[aps,twocolumn]{revtex4}
\usepackage{epsfig}
\usepackage{graphicx}
\begin{document}
\pacs{PACS number(s): 61.43.Hv, 05.45.Df, 05.70.Fh}
\title{Tip Splittings and Phase Transitions in the Dielectric
Breakdown Model: Mapping to the DLA Model.}
\author {Joachim Mathiesen and Mogens H. Jensen}
\affiliation{The Niels Bohr Institute, Blegdamsvej 17, Copenhagen, Denmark.}
\begin{abstract}
We show that the fractal growth described by the dielectric breakdown model 
exhibits a phase
transition in the multifractal spectrum of the growth measure. The
transition takes place because the tip-splitting of branches forms
a fixed angle. This angle is $\eta$ dependent but it
can be rescaled onto an  ``effectively'' universal angle of
the DLA branching process. We derive an analytic rescaling relation which is in agreement
with numerical simulations. The dimension of the clusters decreases linearly
with the angle and the growth becomes non-fractal at an
angle close to $74^\circ$ (which corresponds to $\eta= 4.0\pm 0.3$). 
\end{abstract}
\maketitle

Fractal growth and patterns are common phenomena of nature.
The prototype model for a mathematical description of fractal
growth is the diffusion-limited aggregation (DLA) model \cite{81WS}.
The growth in this model is determined by the electric field
(called the harmonic measure) around the emerging fractal
cluster. The harmonic measure possesses 
multifractal scaling properties and recently, a new insight into
the behavior of this measure was presented by applying the method
of infinitely convoluted conformal mappings \cite{98HL}. In particular it has
been demonstrated that there exists a singular behavior in the
multifractal spectrum and this singularity is a signature of a 
phase transition in thermodynamics formalism of the spectrum \cite{01DJLMP}.
The phase transition in the DLA cluster occurs at a specific moment
$q_c$ of probabilities of the growth. It
has further been demonstrated that the geometrical properties
of the DLA cluster, which gives rise to this transition, is the existence
of a specific branching angle for each new offspring on 
this cluster \cite{01JLMP}. For DLA this critical branching angle was found to be around
27$^o$. Another approach which defines a characteristic
angle in the DLA process, is to consider the stability of a finger
growing in a wedge \cite{98KOOPSS}.

The DLA model has been nicely generalized 
to the dielectric breakdown model (DBM) \cite{84NPW}
where the growth probabilities
at a specific site of the cluster is determined by the value
of the electric field (or the harmonic measure) raised to
a power $\eta$,
i.e. the growth measure at the
interface follows, $\rho_\eta(s)ds\sim |E(s)|^\eta ds$ where DLA
corresponds to the case
$\eta=1$. With varying values of $\eta$ clusters
of different geometry 
are grown each with their own characteristic properties of
the multifractal spectrum. These properties have not been 
outlined before and it is the purpose of this letter to examine
in details possible critical points and phase transitions in the
thermodynamic formalism of the harmonic measure for the
dielectric breakdown model, at varying values of $\eta$. 

The structure of the clusters of the DBM model emerges from an on-going
proliferation and screening, hence stagnation, of branches. In
particular will the protruding branches create fjords in which the
harmonic measure decreases rapidly compared to what happens around the
tips.       
The multifractal properties are best
studied using the recently proposed model of iterated conformal
maps \cite{98HL}, since the deep fjords, numerically, are invisible to the
original approach where random walkers are used as
probes, see \cite{01DJLMP}. 
The model is based on compositions of simple conformal maps
$\phi_{\lambda,\theta}$ which take the exterior of the unit circle 
to its exterior,
except for a little bump at $e^{i\theta}$ of linear size
proportional to $\sqrt\lambda$. We shall here use the mapping
introduced in \cite{98HL} which produces two square root singularities which
we refer to as the branch cuts, and the tip of the bump which we refer to as 
the micro tip. The composition of these mappings is
analog to the aggregation of random walkers in the off-lattice DLA
model. The dynamics is given by 
\begin{equation}
  \label{eq:1}
\Phi^{(n)}(w) = \Phi^{(n-1)}(\phi_{\lambda_{n},\theta_{n}}(w)) \ .
\end{equation}
 
The size of the $n'th$ bump is controlled by the parameter $\lambda_n$
and in order to achieve particles of fixed size we have that, to first
order,
\begin{equation}
  \label{eq:2}
  \lambda_{n} = \frac{\lambda_0}{|{\Phi^{(n-1)}}' (e^{i \theta_n})|^2}. 
\end{equation}
The growth probability $\rho_1(s)$ at the interface of a DLA cluster of size $n$ is, in
the electrostatic picture, proportional to the electric field
\begin{equation}
  \label{eq:4}
  \rho_1(s)ds\sim |E(s)|ds\sim\frac {ds} {|\Phi'|} \ .
\end{equation}
In this case the measure on the unit circle is uniform. When 
the electric field is raised to the power $\eta$, the measure is no
longer uniform,
\begin{equation}
  \label{eq:5}
  \rho_\eta(\theta)d\theta \sim
  \rho_\eta(s(\theta))\left|\frac{ds}{d\theta}\right|d\theta\sim
  \frac{|E|^\eta}{|E|}d\theta\sim|\Phi'(e^{i\theta})|^{1-\eta}d\theta
  \ .
\end{equation}
Numerically we use the
Monte Carlo technique introduced in \cite{01H} in order to choose
$\theta$ according to the distribution $\rho_\eta$. We vary the number of
iterations $T$ such that
for $\eta=1.25$, $T=50$ and for $\eta=4$, $T=400$.


Below we consider both the growth measure and the harmonic
measure. The growth measure is used to derive the multifractal
properties whereas the harmonic measure is used to determine the
physical properties. Let us emphasize that during the growth we always
use the growth measure. First, we consider the growth measure and the
phase transition in the corresponding multifractal spectrum.
 
The moments of the growth probability scales with
characteristic exponents, the generalized dimensions,
\begin{equation}\label{mint}
\int \rho^q(s)ds \sim (1/R)^{(q-1)D_q}\sim n^{(1-q)D_q/D} \ ,
\end{equation}
where $n$ is the number of particles and $R$ the linear size of the cluster. 

Numerically we approximate the integral on the left hand side by the
sum of the field evaluated along the micro tips of the bumps produced
by the bump mappings. The field in DLA will for clusters of size 20000 assume
values below $10^{-20}$ and it follows from (\ref{eq:4}) that it
is impossible with the numerical precision on the unit circle ($\Delta
\theta\simeq 10^{-16})$ to maintain a reasonable resolution in the
physical space $\Delta s= |E|\Delta \theta\simeq10^4$. We therefore
use the resolution increasing
approach introduced in \cite{01DJLMP} where one keeps track on the
dynamics of the branch-cuts.

An easy way to see the existence of a phase transition in the
multifractal spectrum is to look at
the distribution of $\rho_\eta$ sampled along the tips of the
bumps. The distribution will for the smallest values (below some
cutoff value $c$) of $\rho_\eta<c$ scale with a
characteristic exponent $\frac{1-\beta} \beta$. The value of $\beta$ is
calculated by reordering the $N$ computed values of $\rho_\eta$ in
ascending order. In other words, we write them as a sequence
$\{\rho_\eta(i)\}_{i\in I}$ where $I$ is an ordering of the indices
such that $\rho_\eta(i)\leq \rho_\eta(j)$ if $i<j$. We treat the
discrete index $i/N$ as a continuous index $0\le x\le 1$ and therefore $\rho_\eta$ as a non-decreasing function of $x$,
\begin{equation}
\rho_\eta\equiv f(x) \ .
\end{equation}
Numerically we find that the function $f(x)$ obeys a power law with an exponent
$\beta$ for values of $x \ll 1$. From $f(x)$ we calculate the
distribution of $p(\rho_\eta)$ by the transformation formula
\begin{equation}\label{dist}
p(\rho_\eta)\sim\int\delta(\rho_\eta-f(x))dx=\frac 1
{|f'(x(\rho_\eta))|}\sim \rho_\eta^{\frac{1-\beta}\beta} \ .
\end{equation}
With the use of this distribution the moment integral (\ref{mint}) is
rewritten as 
\begin{eqnarray}
\int_0^L \rho_\eta^q ds&=&\int_0^\infty \rho
_\eta^{q}p(\rho_\eta)d\rho_\eta\nonumber\\
&=&k\!\int_0^{c}\rho_\eta^q\rho_\eta^{
  \frac{1-\beta}\beta}d\rho_\eta\! +\!
\int_c^{\infty}\!\!\rho_\eta^q p(\rho_\eta)d\rho_\eta \ ,
\label{plam}
\end{eqnarray}
where $k$ is some normalization constant. The left integral in the final expression
diverges whenever 
\begin{equation}\label{beta}
q\leq q_c=- \frac 1 \beta .
\end{equation}
The phase transition in the multifractal spectrum of the growth
measure takes place for the value $q=q_c$. We shall now argue that
the value of $\beta$, and therefore $q_c$, is independent of the value of $\eta$. In other words the phase transition in the
multifractal spectrum of the growth measure is universal.

\begin{figure}
\centering
\epsfig{width=.45\textwidth,file=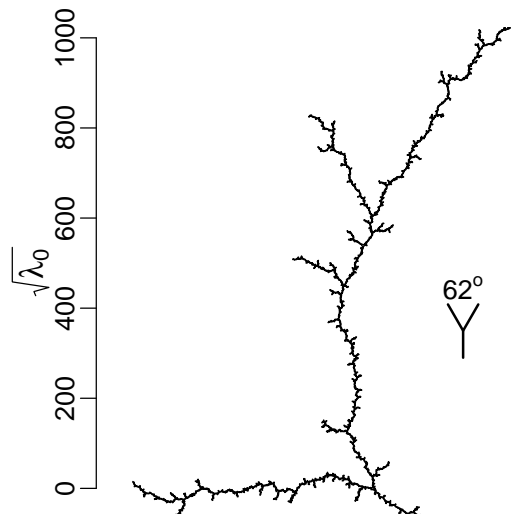}
\caption{Part of a cluster grown with $\eta=3$. The wedge structure at
  the bottom of the fjords is clearly seen and the opening angle
  observed along the aggregate is close to the angle predicted in
  (\ref{ac}) and shown on the figure.}\label{anglefig}
\end{figure}

In \cite{01JLMP} it was shown that the angle defining the branch splitting
deep inside the fjords of DLA was given by a characteristic angle
$\gamma_c(1)=\gamma_c(\eta=1)$. This characteristic angle was also
identified as the reason for the phase transition in the multifractal
spectrum of the harmonic measure of DLA. The electric field along the branches in a wedge with opening angle $\gamma$ scales like
\begin{equation}\label{etascl1}
\rho_1(x)\propto|E(x)|\sim x^{\frac \pi \gamma -1}
\end{equation}      
and therefore in DBM the growth probability inside this
wedge is given by 
\begin{equation}\label{etascl}
\rho_\eta(x)\propto |E(x)|^\eta\sim x^{\eta(\frac \pi \gamma -1)} \ .
\end{equation}
When the exponent $\eta$ is introduced, the growth
probability (\ref{etascl}) inside the wedge is changed and therefore we see a corresponding
change in the geometry, such that e.g. when $\eta>1$ the
field inside a wedge with opening angle $\gamma_c(1)$ is effectively
similar to that of a wedge with a smaller opening angle. 

The wedge structure at the bottom
of the fjords of DLA is not affected by a change in the value of
$\eta$ see Fig. \ref{anglefig}, the physical angle, however, is changed such that the effective angle remains fixed and equal to the angle
observed in DLA. Generally we therefore have for the angle observed physically that $\gamma_c(\eta>1)>\gamma_c(1)$ and $\gamma_c(\eta<1)<\gamma_c(1)$. The angle for a given value of
$\eta$ is found by comparing the exponents in (\ref{etascl}) and (\ref{etascl1})
\begin{equation}\label{ac0}
(\frac{\pi}{\gamma_c(\eta)} -1)\eta =\frac \pi {\gamma_c(1)}-1
\end{equation} 
or
\begin{equation}\label{ac}
\gamma_c(\eta)=\frac{\pi\eta}{\frac{\pi}{\gamma_c(1)}-1 +\eta} \ .
\end{equation}
Fig. \ref{angle} shows together with numerical predictions (see
below) how the critical angle varies with $\eta$. The value of $\gamma_c(1)$ in
(\ref{ac}) and used in the figure is determined numerically from 15 DLA
clusters of size n=20000, $\gamma_c(1)=27^\circ\pm 3^\circ$.

\begin{figure}
\centering
\epsfig{width=.45\textwidth,file=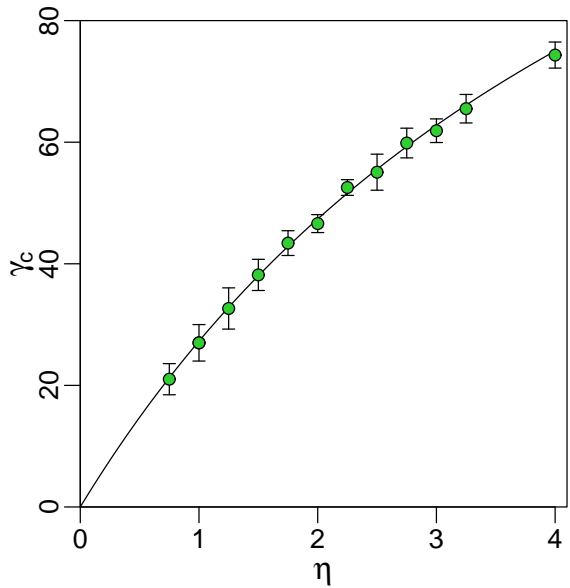}
\caption{The critical angle as function of $\eta$. The line represents
  the critical angle obtained from  the analytical expression (\ref{ac}). Each dot
  represent the numerical average of 4-15 clusters of size $n=18000$. The
  standard deviation of the individual points assumes values in the range between $1.5^\circ$ and $3.4^\circ$.}\label{angle}
\end{figure}

The distribution of $\rho_\eta$ inside a wedge with opening
angle $\gamma$ follows from (\ref{etascl}), with $\alpha=\pi/\gamma$,
\begin{equation}\label{lamwed2}
p(\rho_\eta)\sim\frac{1}{\big(x(\rho_\eta)\big)^{\eta(\alpha-1)-1}}\sim \rho_\eta^{\frac{1-\eta(\alpha-1)}{\eta(\alpha-1)}},
\end{equation}
and if we compare the exponent with that of (\ref{dist}) and insert the
critical angle from (\ref{ac0}) we find that
\begin{equation}
  \label{eq:7}
  \beta=\frac{\pi}{\gamma_c(1)}-1 \ .
\end{equation}
Therefore $\beta$ in expression (\ref{beta}) is independent of $\eta$.

One way, numerically, to calculate the critical angles shown in Fig.
\ref{angle} is first to locate the regions where the distribution in
(\ref{dist}) scales and afterwards perform a direct measurement in
these regions. Such measurements are most likely rather
inaccurate and therefore we turn to (\ref{etascl1}). During the growth
we apply the exponent of $\eta$ as usual but once a cluster is grown
we consider the harmonic measure $\rho_1$ only.
Similarly, as above, we find that inside a wedge
\begin{equation}\label{lamwed}
p(\rho_1)\sim\rho_1^{\frac{1-(\alpha-1)}{\alpha-1}} \ .
\end{equation}
The exponent is compared to the one we calculate numerically from
(\ref{dist}) and from this we find the critical angle as function of
$\beta$. Note that when we consider the harmonic measure and not the
growth measure, $\beta$
is a nontrivial function of $\eta$ and the critical angle is given by 
\begin{equation}\label{gm}
\gamma_c(\eta)=\frac{\pi}{1+\beta(\eta)} \ .
\end{equation}

\begin{figure}
\centering
\epsfig{width=.45\textwidth,file=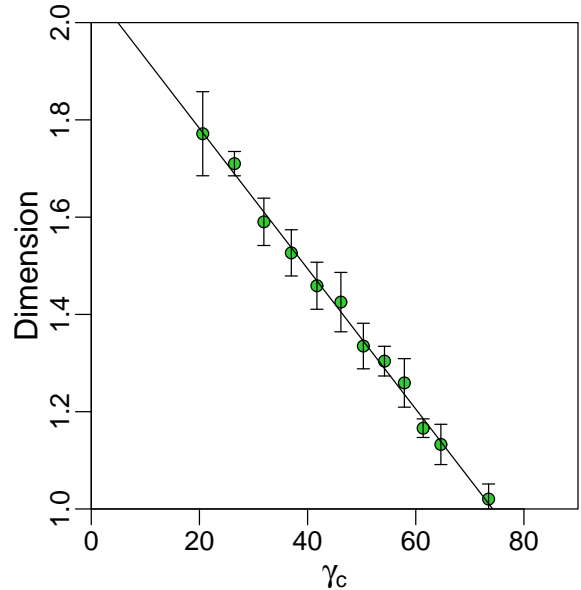}
\caption{The estimated dimension plotted versus the critical angle
  $\gamma_c$, see (\ref{ac}). The range of the critical angle
  corresponds to values of $\eta$ between 0.75 and 3.5. The linear fit
  of the points intersects the line D=1 at an angle $\gamma_c\approx 74^\circ$.}\label{dim}
\end{figure}

Another interesting observation is that the dimension seems to depend
linearly on the critical angle as shown in Fig. \ref{dim}. Note that the linear
fit intersects the line of value equal one before the angle reaches 180
degrees, and therefore the growth is only fractal for $\eta$ below
a finite value. 


The point of intersection is found at the angle $\gamma_c=74^\circ$
and using (\ref{ac}) we find that 
\begin{equation}
  \label{eq:3}
  \eta=4.0\pm 0.3 \ ,
\end{equation}
in agreement with the results obtained in \cite{93SGSHL} and \cite{01H}.
To sum up on these results, we rewrite the relation between the dimension and the
critical angle in terms of the angles $\gamma_2$ and
$\gamma_1$ at which the growth becomes two- and one-dimensional
respectively
\begin{equation}
  \label{eq:6}
  D(\gamma_c)=1+\frac{\gamma_1-\gamma_c}{\gamma_1-\gamma_2}, \quad
  \gamma_2\leq\gamma_c\leq\gamma_1 \ .
\end{equation}
The dimension can also by (\ref{ac}) be written in terms of
$\eta$ and in this case the dependence will no longer be linear but be
of a form similar to that observed in \cite{93SGSHL}. Due to the finite size of the bumps used in the growth we do not
observe, as $\eta\rightarrow 0^+$, that the critical angle of
the fjords vanishes. The bumps will fill up the fjords, and the growth become two dimensional for a non-vanishing value of $\eta$ and
$\gamma_2\approx 5^\circ$. The fill up problem is also the reason why we have
not been able to present data points for values of $\eta$ below
$.75$, because the deep fjords are filled up.

In conclusion, we have studied the critical properties of the
growth of clusters in the dielectric breakdown model. In particular,
we have focused on the branching process and have found that
the branching, on the average, occurs at a fixed angle which
depends on the value of the characteristic parameter $\eta$ 
of the DBM model. The size of the angle in turn determines 
the phase transition point of the growth measure. We have derived 
an analytic expression for the variation of the branching angle
with $\eta$ and found excellent agreement with numerical data. 
Further, we have found a linear dependence of the dimension of
the cluster with the value of critical angle. This linear dependence
results in a prediction of the branching angle at the point where
the growth becomes one-dimensional. It is found to be 
$\gamma_c \approx 74^o$ corresponding to $\eta_c \approx 4$
in agreement with a result obtained in \cite{93SGSHL} and \cite{01H}.
We are in the process of developing a scaling theory for the linear 
dependence of the dimension versus angle.


\end{document}